\documentclass[prb,twocolumn,showpacs,floatfix]{revtex4}
\usepackage{amsmath}
\usepackage{verbatim}
\usepackage{graphicx, epsfig}
\usepackage{amssymb}
\usepackage{bm}

\def\be{\begin{equation}}
\def\ee{\end{equation}}
\def\bea{\begin{eqnarray}}
\def\eea{\end{eqnarray}}
\def\bse{\begin{subequations}}
\def\ese{\end{subequations}}

\def\be{\begin{eqnarray}}
\def\ee{\end{eqnarray}}

\begin{document}

\title{ Boson Hubbard model with weakly coupled Fermions}
\author{Roman M. Lutchyn$^{1,2}$, Sumanta Tewari$^{1}$, and
S. {Das Sarma}$^{1}$}
\author{}
\author{}
\affiliation{$^{1}$Condensed Matter Theory Center, Department of Physics, University of
Maryland, College Park, MD 20742}
\affiliation{$^2$Joint Quantum Institute, Department of Physics,\\ University of Maryland, College Park, MD 20742}
\affiliation{$^3$Department of Physics and Astronomy, Clemson University, Clemson, SC 29634}
\date{\today}

\begin{abstract}
 Using an imaginary-time path integral approach, we develop
 the perturbation theory suited to the boson Hubbard model, and apply it to calculate the effects of a
 dilute gas of spin-polarized fermions weakly interacting with the bosons. The full theory captures both the static and the dynamic effects
 of the fermions
 on the generic superfluid-insulator phase diagram. We find that,
  in a homogenous system described by a single-band boson Hubbard Hamiltonian, the intrinsic
  perturbative effect of the fermions is to generically suppress the insulating lobes and to enhance the superfluid phase.
\end{abstract}

\maketitle

{\it Introduction.}
The boson Hubbard model has long provided the paradigm for studying
one of the simplest quantum phase transitions (QPT), the superfluid
to insulator transition (SIT) in a dilute gas of bosons. Most
satisfactorily, recent experiments~\cite{Greiner_Nature02,
spielman_PRL07} using ultra cold bosonic atoms confined to an
optical lattice, which mimics the boson Hubbard model in a custom
setting,  demonstrated the existence of the SIT in a pristine,
disorder free, boson-only system. By varying the effective
$\frac{t}{U}$ of the ultra cold atoms in optical lattices, where $t$ is the boson nearest-neighbor hopping parameter and $U$ is the on-site
boson-boson repulsion, the
researchers demonstrated the existence of the Mott-insulating (small
$\frac{t}{U}$) and the superfluid (large $\frac{t}{U}$) states in
the time-of-flight experiments.~\cite{Greiner_Nature02,
spielman_PRL07} At some intervening value of $\frac{t}{U}$, then,
there should be a QPT separating the two
states.~\cite{Fisher_PRB89, Sachdev_book}

\begin{figure}
\centering
\includegraphics[width=0.99\linewidth]{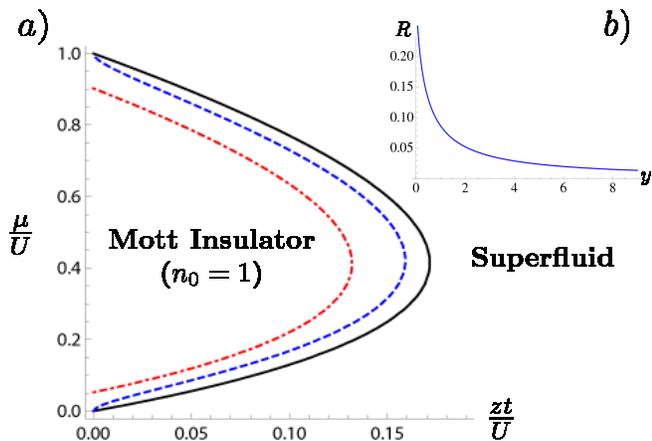}
  \caption{(Color online) a) Main panel: Phase Boundary of the boson Hubbard model with and without the fermions for the boson density $n_0=1$. Solid line describes the
 insulator-superfluid phase boundary without the fermions. The dashed line corresponds to the same phase boundary with the fermions present. The dash-dot line denotes the phase boundary in the static approximation. The regions near the degeneracy points (integer $\mu/U$) are implicitly excluded from this figure~\cite{breakdown}.
Here we used ${U \over 4E_F}\!=\!0.1$ and $\frac{U_{FB}^2}{\Delta
U}\!=\!0.15$. b) Inset: The dependence of the function $R(y)$ on its argument. } \label{fig:phase_boundary}
\end{figure}

An important theoretical question, which has received wide attention~\cite{Lewenstein_PRL04, Buchler_PRB04,
pollet_PRA08, Refael_PRB08, Mathey_PRL04, Sengupta_PRA07, Mering_07}
in light of the recent experiments in the Bose-Fermi
mixtures,~\cite{gunter_PRL06, ospelkaus_PRL06} is what happens to
the insulating and the superfluid phases when fermions are
introduced to the bare boson Hubbard model.
In the case of the bosons weakly interacting with spin-polarized fermions, which are away from half-filling,
this question can be addressed analytically. While some
of the earlier studies~\cite{Buchler_PRB04,
pollet_PRA08} concluded that the region occupied by the superfluid
phase in the phase diagram is enhanced
by fermions, more recent ones~\cite{Refael_PRB08} concluded that the
opposite is true because of an effect akin to the fermionic orthogonality
catastrophe due to the dynamic effects. In this Communication, we address this question by
developing a rigorous perturbation theory suited to the single-band boson
Hubbard model, which captures both the static
and the dynamic effects mediated by the fermions.
Our conclusion is that, in a homogenous, single-band system and in the absence of
loss of cooling due to adding fermions, the fermions
intrinsically shrink the area occupied by the Mott
insulating lobes (Fig.~\ref{fig:phase_boundary}), thus generically enhancing
the superfluid region. The overall effect is qualitatively in the same direction as in the
effects of Ohmic dissipation in enhancing the superconducting phase
coherence in Josephson junction arrays~\cite{Tewari_PRB06} or
in granular superconductors.~\cite{Beloborodov_review} Even though in the current experiments~\cite{gunter_PRL06, ospelkaus_PRL06} the Bose-Fermi interaction strength is not in the perturbative regime, it is possible to tune this coupling and bring it to the perturbative regime~\cite{Ospelkaus_PRL97}. Thus, our predictions can be tested experimentally. Furthermore, in light of our present analytical results (and the results of Refs.~[\onlinecite{Buchler_PRB04,
pollet_PRA08}]) it seems likely that
the observed loss of
superfluid coherence by adding fermions~\cite{gunter_PRL06,
ospelkaus_PRL06} should be attributed to the external
 factors, such as heating~\cite{cramer_PRL08} and self-trapping of
the bosons and fermions.~\cite{luhmann-2007}  Hence, experiments which can avoid such effects (\textit{e.g.}, shallower lattices and lower boson filling factor have reduced boson self trapping due to fermions \cite{luhmann-2007}) are necessary to see the intrinsic effect - enhancement of the superfluidity - due to the fermions.
  We stress that
the perturbation theory of the boson Hubbard model we develop, which deviates from the standard machinery
\cite{Fetter_book} applicable to the free bosons, should have other
important applications, {\it e.g.}, the phase diagram of the boson Hubbard model in the presence of coupling to a dissipative Ohmic bath \cite{Dalidovich} or a second boson species. In general, our method, specifically Eqs.~(\ref{eq:Green's}, \ref{eq:K-def}, \ref{eq:K-calc}), can be taken over in any problem where the Green's function of the boson Hubbard model has to be calculated in perturbation theory.

{\it The model and the results.}
We consider a mixture of bosonic and spin-polarized fermionic atoms
in an optical lattice. The Hamiltonian of the Bose-Fermi system is written as
$H=H_B+H_F+H_{BF}$, with
\begin{align}
&\!H_B\!\!=\!\!\sum_{i}\left(\!\frac{U}{2}
\hat{n}_{i}(\hat{n}_{i}\!-\!1)\!-\!\mu
\hat{n}_{i}\!\right)\!-\!t\sum_{<ij>}\left(b^{\dag}_i b_j
\!+\!H.c.\right)\!,\label{eq:Bose-Hubbard}\\
&H_F\!=\!-t_F\sum_{<ij>}\left(c^{\dag}_i c_j
\!+\!H.c.\right)-\mu_{F}\sum_i c^{\dag}_i c_i,\\
&H_{FB}\!=\!U_{FB}\sum_i \hat{n}_{i} (c^{\dag}_i c_i-n^0_{Fi}).
\label{eq:H}
\end{align}
Here $c^\dag_i$ and $b^{\dagger}_i$ are the fermion and the boson
creation operators on site $i$, $\hat{n}_{i}=b_i^\dag b_i$ is the
boson density operator, $U\!>\!0\, (U_{FB})$ describes the on-site
boson-boson (boson-fermion) interaction, $t(t_F)$ corresponds to the
hopping matrix element for the bosons (fermions), $n^0_{Fi}$ is the
average density of the fermions, and $\mu=\mu_0-U_{FB}n^0_{Fi}$ and
$\mu_F$ are chemical potentials for boson and fermions,
respectively. Here $\mu_0$ is the boson chemical potential without
the fermions.

The partition function of the bare model (without the fermions) can
be written in terms of an imaginary-time path integral over a
complex scalar field $\psi (\mathbf{x}, \tau)$,~\cite{Fisher_PRB89,
Sachdev_book} where $\tau$ is the imaginary time. The action in
terms of $\psi(\mathbf{x}, \tau)$ takes the form of a
$\phi^4$ theory, see Eq.~(\ref{eq:Phi4}). In this
description, the details of the bare Hamiltonian are hidden in the
coefficients of the various terms of the action. For example, the
coefficient, $r$, of the term $|\psi(\mathbf{x},\tau)|^2$ (see
below) is determined by the Green's function, $\langle
T_{\tau}b_i(\tau)b_i^{\dagger}(0)\rangle$, of the bosons~\cite{Fetter_book}, where $\langle
...\rangle$ denotes average with respect to the on-site part of the
boson Hubbard Hamiltonian. In mean field theory, $r=0$ gives the
locus of the insulator ($r > 0, \langle
\psi(\mathbf{r},\tau)\rangle =0$) to the superfluid ($r < 0, \langle
\psi(\mathbf{r},\tau)\rangle \neq 0$) QPT, revealing the Mott
insulating lobes in the phase diagram.~\cite{Fisher_PRB89, Sachdev_book}

With fermions, a similar description of
the partition function still holds, but now the boson Green's
function must incorporate the perturbative effects of the
boson-boson interaction \textit{mediated} by the fermions.
We stress that this mediated interaction is manifestly non-local in
both space and time. Therefore, it is not obvious that this problem can be treated in an effective Weiss-like single-site theory as done in Ref.~[\onlinecite{Refael_PRB08}]. The perturbative
corrections to the boson Green's function cannot be calculated by using the standard
diagrammatic machinery \cite{Fetter_book} either, because the bare
Hamiltonian is an interacting one and the interaction $U$ has to be
treated non-perturbatively. We solve this problem by noting that
we can still calculate the needed correlation functions exactly by making use of the eigenstates of the
number operators $\{n_i\}$. A modified linked-cluster theorem still
holds which gets rid of all the divergences encountered in the
perturbation theory. The locus of the equation, $r^{\prime}=0$,
where $r^{\prime}$ includes the perturbative corrections to the boson Green's
function, provides the phase boundary between
the superfluid and the insulating states. Our central result for the
phase boundary is shown in Fig.1. Below we
give a summary of the methods and the calculations used to arrive at
the results. The details of the calculations will be given
elsewhere.~\cite{forthcoming}

{\it Summary of the methods.} To the lowest order in $U_{FB}$, the effect of the fermions on the
constituent bosons is a trivial shift of the boson chemical
potential $\mu=\mu_0-U_{FB}n^0_{Fi}$. All the non-trivial effects
appear in the second order in $U_{FB}$.
By integrating out the fermions,
the imaginary-time partition function becomes
(we assume here zero temperature
$T\rightarrow 0$)
\begin{align}
Z&=\int \mathfrak{D} b_i^* \mathfrak{D} b_i \exp\left(-S_{\rm eff}[ b_i^*, b_i]\right) \label{eq:part}\\
S_{\rm eff}[b_i^*,b_i]&=\int_{0}^{\beta} d\tau \left(\sum_i b^*_i \partial_{\tau} b_i+H_B \right)\label{eq:eff_action}\\
&-\sum_{ij} \int_0^{\beta} d \tau_1 \int_0^{\beta} d \tau_2
n_{i}(\tau_1)M_{ij}(\tau_1\!-\!\tau_2) n_{j}(\tau_2).\nonumber
\end{align}
In the second order in $U_{FB}$, the integral over the fermion degrees of freedom gives rise to an effective non-local density-density interaction for the bosons with the function $M_{ij}(\tau_1-\tau_2)$ being,
\begin{equation}
M_{ij}(\tau_1-\tau_2)=\frac{U_{FB}^2}{2}\left\langle \Delta n_{Fi}(\tau_1) \Delta n_{Fj}(\tau_2) \right\rangle.\label{eq:M(ij)}
\end{equation}
In the frequency and momentum domain, $M_{\bm q}(\Omega_n)$ is proportional to the fermion polarization function, and in 2D is given by,
\begin{equation}\label{eq:polarization}
M_{\bm q}(\Omega_n)=\frac{U_{FB}^2}{2\Delta} \left(1-\frac{|\nu_n|}{\sqrt{\nu_n^2+ k^2}}\right).
\end{equation}
 Here, $\nu_n=\Omega_n/4E_F$ and $k=q/2k_F$, with $E_F$ and $k_F$ being the Fermi energy and the Fermi momentum, respectively. $\Delta$ is the fermion mean level-spacing, $\Delta=1/\nu_FV$, with $\nu_F$ the density of states at the Fermi level and $V$ the volume of the unit cell. Equation~(\ref{eq:polarization}) is valid for $k<1$. Here, for simplicity, we consider a 2D system. However, our qualitative conclusions hold for the 3D case as well \cite{forthcoming}.

Using the Hubbard-Stratonovich transformation with a complex scalar field $\psi_i(\tau)$, we integrate out the bosonic fields to write,
$Z=Z_0 \int \mathfrak{D} \psi_i  \mathfrak{D} \psi^*_i \exp(-S[\psi_i,\psi_i^*])$, where the action $S[\psi_i, \psi_i^*]$ is given by,
\begin{align}\label{Spsi}
S[\psi_i, \psi_i^*]&=\int_0^{\beta} d \tau \sum_{i,j} \psi^*_i(\tau)
w^{-1}_{ij} \psi_j(\tau)\\\nonumber &- \ln \! \left  \langle \exp
\left[ \int_0^{\beta} d \tau \sum_i b_i(\tau)
\psi_i^*(\tau)+ H. c. \right] \! \right \rangle \!.
\end{align}
Here the matrix elements of the symmetric matrix, $w_{ij}$, are
equal to $t$ for the nearest neighbors and zero otherwise. The
expectation value in Eq.~(\ref{Spsi}) is taken with respect to the action $S_{\rm eff}[b_i^{*},b_i]$ (with $t=0$). By expanding $S[\psi, \psi^*]$ up to the fourth power of the field $\psi$, and taking the continuum limit, we arrive at the action of an effective complex $\psi^4$ field theory,
\begin{equation}
\!\!{\cal{S}}[\psi,\!\psi^*]\!\!=\!\!\!\int \!\! d\bm x  \!\! \left(\!\!c_{1} \psi^* \frac{\partial\psi}{\partial
\tau}+\!c_{2}\! \left|\frac{\partial\psi}{\partial
\tau}\right|^2\!\!+\! c\! \left|\bm \nabla{\psi}\right|^2\!+\!r\! \left|\psi\right|^2\!+\!u\!
\left|\psi\right|^4\!\right)\!\! \label{eq:Phi4}
\end{equation}
with $\bm x=\{\bm r,\tau \}$. The coupling constants $c_{1},c_2, c, r, u$ are given by the correlation functions of the boson Hubbard model with $t=0$.
In mean field theory, the phase boundary between the
superfluid and insulating states can be obtained by
setting the coefficient $r$ to zero:
\begin{align}\label{eq:r}
r \propto \frac{1}{z t}+\int_{-\beta}^{\beta} d\tau {\bf G}_i(\tau)=0,
\end{align}
where ${\bf G}_i(\tau)=-\langle T_{\tau} b_i(\tau)b_i^{\dag}(0)\rangle$ is the single-site boson Green's function, which, in the presence of the fermions, should include the effective fermion-mediated density-density interaction.
Without the fermions, this Green's function is given by, \cite{Sachdev_book}
\begin{equation}
G_i(i\omega_n)\!=\!\!\left[\frac{(n_0\!+\!1)}{i\omega_n\!-\!\delta E_p}\!-\!\frac{n_0}{i\omega_n\!+\!\delta E_h}\right],
\label{bare-green}
\end{equation}
where $\delta E_p$ and $\delta E_h$ are particle and hole excitation energies: $ \delta E_p={U n_0 -\mu}$ and $\delta
E_h={\mu -U (n_0 -1)}$, and $n_0$ is the number of bosons per site minimizing the ground state energy. Thus, the problem is now reduced to the calculation of
the on-site full boson Green's function by computing the corrections
to Eq.~(\ref{bare-green}). As we show below, this can be done
perturbatively in $U_{FB}$.

The calculation of the perturbative corrections to the boson Green's
function is non-trivial because the bare Hamiltonian, $H_B$ (with
$t=0$), is not quadratic in the boson operators. Therefore, one
cannot use the standard diagrammatic techniques, \cite{Fetter_book}
because the Wick's theorem does not hold. To make progress, we write
the corrections to the Green's function using the cumulant
expansion:
\begin{align}\label{eq:Green's}
\! \langle \! \langle T_{\tau}
b_i(\tau)b^{\dag}_i(0) \rangle \! \rangle\!&\!=\!\!\langle T_{\tau}
b_i(\tau)b^{\dag}_i(0) \rangle\\
\!&\!+\!\!\sum_{jl}\!\int_0^{\beta} \!\! d \tau_1\!\! \int_0^{\beta} d \tau_2
M_{jl}(\tau_1\!-\!\tau_2)K_{ijl}(\tau,\tau_1,\tau_2),\nonumber
\end{align}
where $\langle\!\langle ...\rangle\!\rangle$ denotes the Green's function which includes the perturbative corrections. In the Mott-insulating state, it is convenient to calculate the correlation function $K_{ijl}(\tau,\tau_1,\tau_2)$ in the second quantized representation:
\begin{align}\label{eq:K-def}
K_{ijl}(\tau,\tau_1,\tau_2)=&\langle T_{\tau}
b_i(\tau)b^{\dag}_i(0) n_{j}(\tau_1) n_{l}(\tau_2) \rangle\\
-&\langle T_{\tau}
b_i(\tau)b^{\dag}_i(0) \rangle \langle  T_{\tau} n_{j}(\tau_1) n_{l}(\tau_2) \rangle. \nonumber
\end{align}
Given that the on-site part of the boson Hubbard Hamiltonian conserves
the number of bosons, the correlation functions above can be
calculated exactly using the particle-number eigenstates.~\cite{forthcoming}  The terms in $K_{ijl}(\tau,\tau_1,\tau_2)$ contributing to static and dynamic screening are given by,
\begin{align}\label{eq:K-calc}
\! & K_{ijl}(\tau,\tau_1,\tau_2)
\!=
\Theta(\tau)\Theta(\tau_1)\Theta(\tau_2)\Theta(\tau\!-\!\tau_1)\Theta(\tau\!-\!\tau_2)\nonumber\\
& \times
\left[(\delta_{ij}\!+\!\delta_{il})n_0(n_0\!+\!1)\!+\!\delta_{ij}\delta_{il}(n_0\!+\!1)\right]\exp \left(-\delta
E_p\tau\right)\nonumber\\
& +
\Theta(-\tau)\Theta(-\tau_1)\Theta(-\tau_2)\Theta(\tau_1-\tau)\Theta(\tau_2-\tau)\nonumber\\
& \times \left[-(\delta_{ij}+\delta_{il})n_0^2+\delta_{ij}\delta_{il}n_0\right]\exp\left(\delta
E_h\tau\right)+....
\end{align}
It is important to note that $K_{ijl}(\tau, \tau_1, \tau_2)$ is irreducible and cannot be factored into the product of the bare Green's functions, as would have been possible if Wick's theorem were applicable.

We now proceed to calculate the effects of the fermions by first
approximating $M_{\bm q}(\Omega_n)$ in Eq.~(\ref{eq:polarization})
by the constant piece, $M_{\bm
q}(\Omega_n)\!\sim\!\frac{U_{FB}^2}{2\Delta}$ (static
approximation, see also Ref.~[\onlinecite{Buchler_PRB04, Refael_PRB08}]). By substituting
the corresponding expression for $M_{jl}$,
$M_{jl}(\tau_1\!-\!\tau_2)\!=\!\frac{U_{FB}^2}{2\Delta}\delta_{lj}\delta(\tau_1-\tau_2)$,
into Eq.~(\ref{eq:Green's}), and carrying out the imaginary-time
integrals, we find the following expression for the Green's function
at zero frequency~\cite{breakdown},
\begin{align}\label{eq:Green's_stat}
\!{\bf G}_i(0)\!=\! - \frac{n_0\!+\!1}{\delta E_p}\!\left[\!1\!+\!\frac{U_{FB}^2\!(1\!+\!2n_0)\!}{2 \Delta \delta E_p}\!\right]\!-\!\frac{n_0}{\delta E_h}\!\left[\!1\!+\!\frac{U_{FB}^2\! (1\!-\!2n_0)\!}{2 \Delta \delta E_h}\!\right]\!.
\end{align}
Alternatively, we could substitute the static, on-site form of
$M_{ij}(\tau_1-\tau_2)$ directly into the action,
Eq.~(\ref{eq:eff_action}), and calculate the Green's function
exactly. It is easy to see that, in the static approximation, the mobile fermions simply renormalize $\mu$ and $U$ of the bare boson Hubbard Hamiltonian $H_B$: $U\rightarrow U- U_{FB}^2/\Delta$ and $\mu\rightarrow \mu+U_{FB}^2/2\Delta$. The exact Green's function, thus, can simply be obtained by substituting these renormalized parameters in Eq.~(\ref{bare-green}).
After expanding the result to the second order in $U_{FB}$, the resulting expression exactly matches \cite{forthcoming} that in Eq.~(\ref{eq:Green's_stat}). This
validates the correctness of our perturbation theory. Using Eq.~(\ref{eq:r}) one can see that, in the static approximation,
the fermions markedly shrink the area of the Mott-insulating lobes in the phase diagram (see Fig.~\ref{fig:phase_boundary}).

The static screening approximation for $M_{\bm q}(\Omega_n)$ does not, however, take into account the important retardation effects \cite{Refael_PRB08} and the spatially non-local nature of the interaction kernel in Eq.~(\ref{eq:M(ij)}). By substituting the full expression for $M_{ij}(\tau_1-\tau_2)$ into Eq.~(\ref{eq:Green's}), and doing the imaginary-time integrals as well as carrying out the summation over $j$ and $l$, we obtain the following expression for the boson Green's function at zero frequency,
\begin{align}\label{eq:Green's_dyn}
&{\bf G}_i(0)=\\
&\!-\! \frac{n_0\!+\!1}{\delta E_p}\!\left[1\!+\!\frac{U_{FB}^2}{\Delta \delta E_p}R\!\left(\!\frac{\delta E_p}{4E_F}\!\right)\!\right]\!-\!\frac{n_0}{\delta E_h}\!\left[1\!+\!\frac{U_{FB}^2}{\Delta \delta E_h}R\!\left(\!\frac{\delta E_h}{4E_F}\!\right)\!\right]\!.\!\nonumber
\end{align}
Here we introduced the dimensionless function $R(y)$:
\begin{align}\label{eq:Rz}
R(y)\!&=\! \frac{4}{\pi^2}\int_{0}^{1} kdk\!\int_{0}^{\infty} d \nu
\left[1-\frac{|\nu|}{\sqrt{k^2+\nu^2}}\right]
\frac{y}{\nu^2+y^2}\\
&=\frac{4}{\pi^2}\left[\frac{\pi}{4}\!+\!y  - \frac{\pi}{2} y^2 \!+\! y \sqrt{y^2\!-\!1}
\sec^{-1}(y)\right] \nonumber.
\end{align}
The inset in Fig.~\ref{fig:phase_boundary} depicts the behavior of the monotonic function $R(y)$ as a function of its argument.
 As follows from Eq.~(\ref{eq:Green's_dyn}), the importance of the fermion renormalization effects is determined by the ratio of $\delta E_{p/h}$ and $E_F$.
 When the fermion density is small, \emph{i.e.}, $\delta E_{p/h}/E_F\!\gg\! 1$, the corrections to the Green's function are suppressed since $R (y\!\gg\! 1)\!\rightarrow \!0$. In the opposite limit, $\delta E_{p/h}/ E_F\ll 1$, the function $R(y\ll 1)\!\sim \!1$, and thus, for a given value of $U_{FB}$, the effects of the fermions on the bosons are more pronounced. Finally, using Eq.~(\ref{eq:Green's_dyn}) and Eq.~(\ref{eq:r}), we calculate the phase diagram on the $(\mu-t)$ plane as shown in Fig.~\ref{fig:phase_boundary}. We emphasize that the net effect of the fermions is to suppress the Mott-insulating lobes and enhance the superfluidity.

 The above result is consistent with numerical calculation of Ref.~[\onlinecite{pollet_PRA08}] and is in disagreement with the conclusions of Ref.~[\onlinecite{Refael_PRB08}]. We note that the correctness of our formalism for the perturbative evaluation of the Green's function (in the static screening approximation) was confirmed independently, see the discussion after Eq.~(\ref{eq:Green's_stat}). The generalization of the scheme to the dynamical screening is straightforward and amounts to only taking the frequency and momentum integrals, mandated by Eq.~(\ref{eq:Green's}). Thus, we are able to calculate the perturbative effects to the boson Hubbard model of an arbitrary
  time- and space-dependent interaction kernel. In contrast, it is not obvious that a spatially non-local interaction kernel, such as that in Eq.~(\ref{eq:M(ij)}), can be properly treated in the Weiss-like self-consistent mean-field theory employed in Ref.~[\onlinecite{Refael_PRB08}]. Note also that the function $M_{\bm q}(\Omega_n)$ is positive definite for all momenta and frequencies. Therefore, the net effect of the full interaction kernel is qualitatively similar to its constant piece (the static approximation), even though the latter significantly overestimates the suppression of the insulating phase. Thus, our qualitative conclusions should be valid for 3D systems as well. \cite{forthcoming} We note that the sign of the phase boundary shift can be predicted from the sign of the fermion density-density correlation function, while the magnitude of the corrections to the phase diagram depends on the microscopic details such as the ratio of $\delta
  E_{p/h}$  and $E_F$ as follows from Eqs.~(\ref{eq:Green's_dyn}) and (\ref{eq:Rz}). Finally, we emphasize that, near the degeneracy points, where the excitation energy $\delta
  E_{p/h}$ is smaller than $U_{FB}^2/\Delta$, our perturbation theory breaks down, see Eqs.~(\ref{eq:Green's_stat}) and~(\ref{eq:Green's_dyn}).
  Thus, the effect of fermions on the boson Hubbard phase diagram near these points is an open question.

{\it Conclusion.}
In summary, we develop a framework for carrying out the perturbation theory for the boson Hubbard model, and use it to calculate the effects of a dilute gas of spin-polarized fermions weakly interacting with the bosons. The full theory captures both the static and the important dynamic effects of the fermions on the constituent bosons.
We find that within single-band boson Hubbard model the net effect of the fermions is to inherently suppress the Mott-insulating lobes and enhance the superfluid phase in the generic Bose-Hubbard phase diagram.

 We thank K.~Yang, G.~ Refael, E.~Demler, T.~Porto, T.~Stanescu, E.~Hwang, and C. W. Zhang  for stimulating discussions. This work is supported by ARO-DARPA.


\begin{thebibliography}{17}
\bibitem{breakdown} Near the degeneracy points~[\onlinecite{Sachdev_book}] where $\delta
  E_{p/h}$ is of the order~of $U_{FB}^2/\Delta$ the perturbation theory breaks down.
\end{thebibliography}
\end{document}